\documentclass[letter, superscriptaddress, prl, twocolumn, nofootinbib]{revtex4}

\usepackage{verbatim, amsmath, graphicx}

\begin{document}

\title{Unstable growth of curvature perturbation in non-singular bouncing cosmologies}

\author{BingKan Xue}
\affiliation{Department of Physics, Princeton University, Princeton,
New Jersey 08544, USA}

\author{Paul J. Steinhardt}
\affiliation{Department of Physics, Princeton University, Princeton,
New Jersey 08544, USA}
\affiliation{Princeton Center for Theoretical
Physics, Princeton University, Princeton, New Jersey 08544, USA}

\begin{abstract}
We consider non-singular bouncing cosmologies, such as the new ekpyrotic model, in which the universe undergoes a slow contraction phase with equation of state $w \gg 1$, followed by a bounce that occurs at a finite scale factor when quantum gravity corrections are still negligible. Such a non-singular bounce requires a violation of the null energy condition in which $w$ falls below $-1$ at some time before the bounce. In this paper, we show that a component of the adiabatic curvature perturbations, though decaying and negligible during the ekpyrotic phase, is exponentially amplified just before $w$ approaches $-1$, enough to spoil the scale-invariant perturbation spectrum. We discuss how the problem may be avoided, for example, in singular bounces.
\end{abstract}

\maketitle

As an alternative to inflationary cosmology, the ekpyrotic scenario \cite{Khoury:2001wf, Lehners:2008vx} resolves the homogeneity, flatness and isotropy problems of standard cosmology by having the universe undergo a period of ultra-slow contraction followed by a bounce to the standard expanding phase. Ultra-slow contraction (\emph{ekpyrotic phase}) occurs when the dominant energy component has an equation of state $w \gg 1$. During contraction, this component grows faster than all other contributions to the cosmic expansion, including the spatial curvature and anisotropy, and thereby drives the universe into an exponentially flat and isotropic state leading up to the bounce \cite{Erickson:2003zm}.

As for the bounce, two possibilities have been discussed. In ``singular bounce'' models, such as the cyclic model \cite{Khoury:2001bz}, the universe contracts towards a ``big crunch'' until the scale factor $a(t)$ is so small that quantum gravity effects become important. The presumption is that these quantum gravity effects introduce deviations from conventional general relativity and produce a bounce that preserves the smooth, flat conditions achieved during the ultra-slow contraction phase.

The second possibility is a ``non-singular bounce'', such as in the ``new ekpyrotic model'' \cite{Buchbinder:2007ad}, where the universe stops contraction and reverses to expansion at a finite value of $a(t)$ where classical general relativity is still valid. A significant advantage of this scenario is that the entire cosmological history can be described by 4d effective field theory and classical general relativity, without invoking extra dimensions or quantum gravity effects. However, for the bounce to happen within classical general relativity, $w$ must fall below $-1$ for a sustained period, \emph{i.e.} a violation of the null energy condition (NEC). The NEC violating energy component is commonly chosen to be a scalar field that undergoes \emph{ghost condensation} \cite{ArkaniHamed:2003uy,note}.

In this paper, we show that the non-singular bounce creates a serious problem for cosmological perturbations. Observations
of the cosmic microwave background (CMB) and large-scale structure have provided evidence for a scale-invariant perturbation spectrum. However, in the ekpyrotic model when the scale-invariant curvature perturbation is generated during or just after the ekpyrotic phase, a potentially dangerous component of adiabatic curvature perturbations is created at the same time.  This mode has been previously ignored because, after exiting horizon when $w \geq 1$, its amplitude becomes exponentially suppressed on large length scales compared to the scale-invariant modes. In a singular bounce, this mode remains completely negligible because $w\geq 1$ all the way up to the bounce. However, for the non-singular bounce, the ekpyrotic phase must end and $w$ must fall below $-1$. We show that, right before crossing $w=-1$, the adiabatic mode undergoes exponential amplification such that the scale-invariant spectrum is spoiled and perturbation theory breaks down.

\begin{figure}
\includegraphics[width=0.4\textwidth]{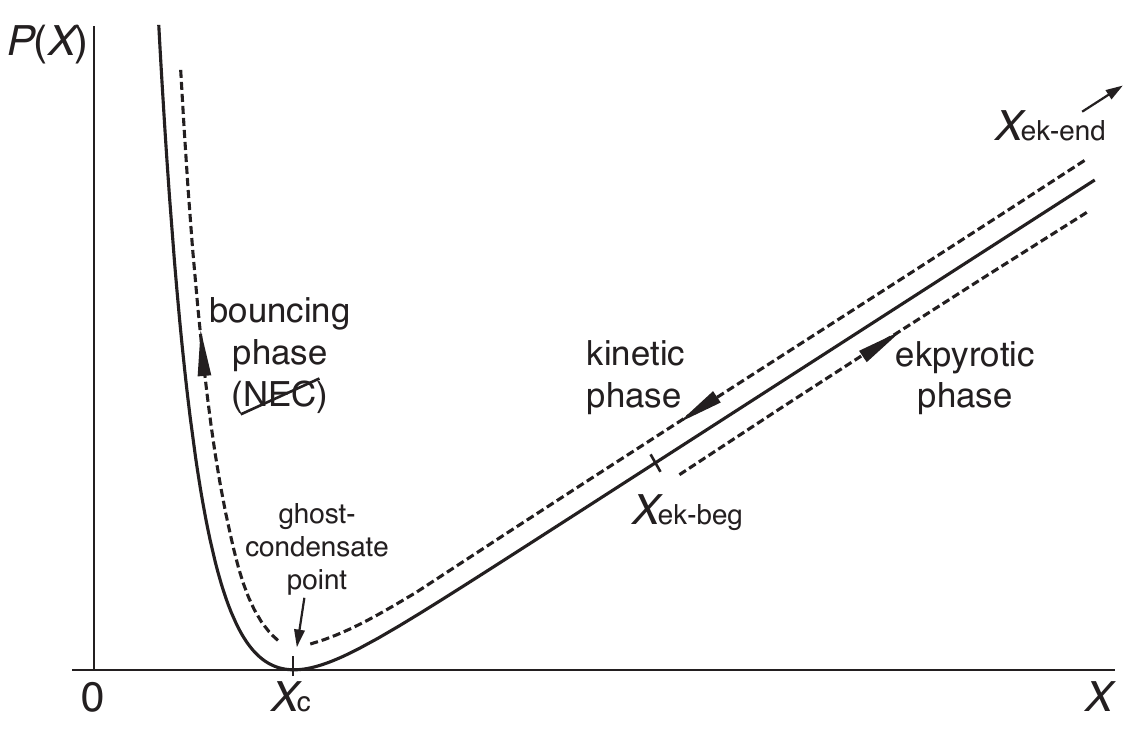}
\caption{The kinetic term $P(X)$ versus $X \equiv \frac{1}{2} (\partial \phi)^2$ for the ghost-condensate scalar field $\phi$. As indicated by the dotted lines with arrows, during the ekpyrotic phase ($X > X_\text{c}$), the approximately canonical kinetic term $P(X) \approx X$ grows exponentially by a factor of $e^{2 N_\text{tot}}$; After the ekpyrotic phase, $X$ decreases by an even greater factor to reach $X=X_\text{c}$. This exponential decrease in $X$ is directly related to the problem with non-singular bounces.}
\label{fig:kinetic}
\end{figure}

\begin{figure}
\includegraphics[width=0.4\textwidth]{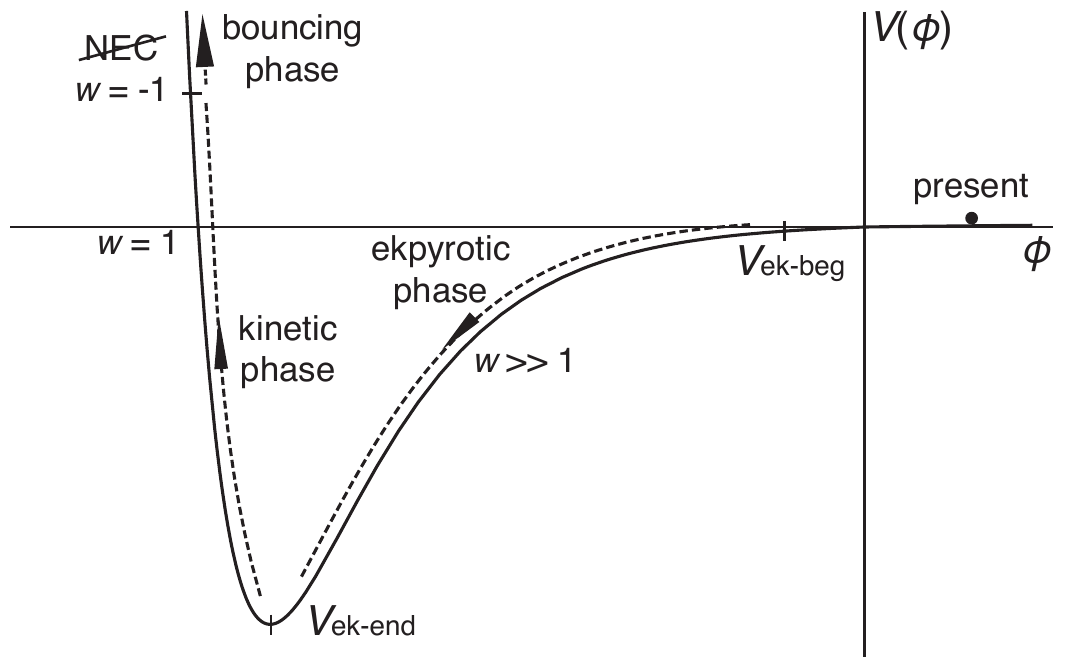}
\caption{The effective potential $V(\phi)$ in the ekpyrotic model with a non-singular bounce.  The evolution along the potential during the ekpyrotic and bounce phases is from right to left: the ekpyrotic phase refers to the exponential decline from $V_\text{ek-beg}$ to $V_\text{ek-end}$ near the minimum of the potential; then there is a brief kinetic energy dominated phase as $V$ quickly rises; and finally the non-singular bouncing phase occurs as $V$ climbs sufficiently above zero.}
\label{fig:potential}
\end{figure}

For our study we take the new ekpyrotic model \cite{Buchbinder:2007ad} as a generic example of non-singular bounces. In this model both the ekpyrotic phase and the ghost-condensate phase are driven by the same scalar field, since otherwise one encounters the problem that the energy of the ghost-condensate field is so exponentially diluted away after the ekpyrotic phase that it is too insignificant to cause the bounce. This framework can be described by an effective Lagrangian
\begin{equation}
\mathcal{L} = \sqrt{-g} \left( \tfrac{1}{2} R + P(X) - V(\phi)\right), \quad X
\equiv - \tfrac{1}{2} \left( \partial \phi \right) ^2,
\end{equation}
assuming a FRW background metric with signature $(-+++)$, and reduced Planck units $8 \pi G \equiv 1$. 
The kinetic term $P(X)$ is designed as in Fig.~\ref{fig:kinetic},
where it is linear for large $X$, $P(X) \approx X$, but has a minimum at a low energy scale $X_\text{c}$. The potential $V(\phi)$ is sketched in Fig.~\ref{fig:potential},
where, beginning from the right hand side, $V$ is approximated by a negative exponential $-V_0 \, e^{- \sqrt{2/p} \, \phi}$ over a range between $V_\text{ek-beg}$ and $V_\text{ek-end}$, then bottoms out and undergoes a steep rise.  The universe evolves through the ekpyrotic, kinetic and bouncing phases, as indicated in the figures.

During the ekpyrotic phase, the Friedmann equation has an attractor solution,
\begin{align} \label{eq:attractor}
& a = \big( \tfrac{-t}{-t_\text{ek-end}} \big)^p, 
\quad H = - \tfrac{p}{(-t)}, \\
& \phi = \sqrt{2p} \, \log \Big( \sqrt{\tfrac{V_0}{p(1-3p)}} \, (-t) \Big), \nonumber
\end{align}
where $t$ is negative and increasing towards zero, and we normalize $a_\text{ek-end} = 1$.
This solution has a constant equation of state, $w = \frac{2}{3p} - 1 \gg 1$, if we choose $p \ll 1$. 
The potential energy and the nearly canonical kinetic energy satisfy the scaling relation  
\begin{equation} \label{eq:scaling}
\tfrac{3}{2} H^2 = \tfrac{1}{1-w} V = \tfrac{1}{1+w} X, \quad X = \tfrac{1}{2} \dot{\phi}^2.
\end{equation}
Therefore, there is an exponential suppression of any initial curvature and anisotropy, determined by the ratio of the kinetic energy densities at the end and beginning of the ekpyrotic phase,
\begin{equation} \label{eq:efolds}
e^{2 N_{\text{tot}}} \equiv \frac{(H^2)_\text{ek-end}}{(H^2)_\text{ek-beg}}
= \frac{X_\text{ek-end}}{X_\text{ek-beg}}.
\end{equation}

A brief kinetic energy dominated phase follows after the field reaches the bottom of the potential and rises towards $V=0$. During this phase the equation of state $w$ rapidly decreases according to
\begin{equation} \label{eq:wdot}
\dot{w} = 2 \sqrt{3(1+w)} \, H \left[ \Big( \tfrac{-V_{,\phi}}{3 H^2} \Big) + \tfrac{w-1}{2} \sqrt{3(1+w)} \right].
\end{equation}
The first term in the brackets represents the ratio of the gradient force to the total energy, which is a large positive factor during the sharp rise in the potential. Therefore $\dot{w} < 0$ in this phase, since $H$ is negative.

In singular bounces \cite{Khoury:2001bz}, the potential approaches zero from below as $\phi \rightarrow -\infty$, so that $w$ remains $\geq 1$ until the bounce. In contrast, for a non-singular bounce, the potential rises above zero as in Fig.~\ref{fig:potential}, so that, as the field climbs, $w$ decreases below 1 and eventually crosses $-1$ some time before the bounce. This stage occurs in much less than one Hubble time, during which $H$ and $a$ are nearly constant. The kinetic term in the Lagrangian remains canonical for $X \gtrsim X_\text{c}$.

The bouncing phase begins when the field climbs sufficiently far up the potential that $X$ falls below $X_\text{c}$ and the ghost condensate phase  initiates. In this phase, $\dot{H} = - X P_{,X} = - \frac{1}{2} \rho ( 1 + w )$ becomes positive and $w$ falls below $-1$, violating the NEC. The non-singular bounce occurs as contraction slows and ultimately halts at a finite value of $a$.

For consistency, a necessary condition is (Fig.~\ref{fig:kinetic})
\begin{equation} \label{eq:consistence}
X_{\text{ek-end}} \gg X_{\text{ek-beg}} \gg X_\text{c}, 
\end{equation}
which keeps the kinetic term $P(X)$ canonical throughout the ekpyrotic phase.
The speed of sound is
\begin{equation} \label{eq:cs2}
c_s^2 = \frac{P_{,X}}{2X P_{,X} - P} \approx 1, 
\end{equation}
when $P(X)$ is linear; in the bouncing phase when $w < -1$,
$c_s^2$ becomes small and negative, but it does not cause instabilities if the universe bounces and exits the ghost-condensate
phase within a Hubble time or so \cite{Buchbinder:2007ad, Creminelli:2006xe}.

The serious problem appears when we consider perturbations around the background evolution.
Here we focus on the widely studied gauge-invariant variable, the comoving curvature perturbation $\mathcal{R}$.
Perturbations in other gauges will be discussed in \cite{Xue}, where we show that results agree at the bounce, so that the initial conditions for the expanding phase are consistent in all gauges.
Different Fourier modes $\mathcal{R}_k$ of the comoving curvature perturbation, labeled by the comoving wave number $k$, obey the equation of motion \cite{Lehners:2008vx}
\begin{equation}
\label{eq:R}
\mathcal{R}_k^{\prime \prime} + 2 \frac{z^\prime}{z} \mathcal{R}_k^{\prime} + c_s^2 k^2 \mathcal{R}_k = 0,
\end{equation}
where $z = a \sqrt{\tfrac{-\dot{H}}{c_s^2 H^2}}$, and prime denotes derivative with respect to
conformal time $\tau$, $\frac{d}{d \tau} = a \frac{d}{dt}$. For small $k$, the equation is formally solved in
expansions of $k^2$,
\begin{equation}
\mathcal{R}_k = \mathcal{R}_k^{(0)} - k^2 \int \frac{d \tau}{z^2} \int d\tau c_s^2 z^2 \mathcal{R}_k,
\end{equation}
where the leading order term is the sum of a constant term $\mathcal{R}_k^\text{const}$ and an integral term $\mathcal{R}_k^\text{int}$
\begin{equation} \label{eq:R0}
\mathcal{R}_k^{(0)} = \frac{C_1}{\sqrt{k}} + \frac{C_2}{\sqrt{k}}
\int_0^t \frac{k dt}{a} \frac{c_s^2 H^2}{a^2 (- \dot{H})} \equiv \mathcal{R}_k^\text{const} + \mathcal{R}_k^\text{int}.
\end{equation}
The dimensionless constants $C_1$ and $C_2$ are numbers of order $O(1)$, as can be found by matching to Minkowski vacuum condition at the beginning of the ekpyrotic phase when the mode is deep inside the horizon.

During the ekpyrotic phase, $\frac{-\dot{H}}{H^2} = \frac{3}{2} (1+w) = \frac{1}{p}$
is nearly constant and $a \sim (-t)^p$ does not change significantly.
Therefore, the integral term decreases as $\mathcal{R}_k^\text{int} \sim k(-t)$ as $t \rightarrow 0^-$. It is comparable to $\mathcal{R}_k^\text{const}$ at horizon crossing ($k \sim a H$), but becomes exponentially small by the end of the ekpyrotic phase,
\begin{equation} \label{eq:Nk}
\left| \frac{\mathcal{R}_k^\text{int}}{\mathcal{R}_k^\text{const}} \right| _\text{ek-end}
\approx \frac{p \, k}{(aH)_{\text{ek-end}}}
\approx \sqrt{\frac{X_k}{X_{\text{ek-end}}}}\equiv  \, e^{-N_k} \ll 1,
\end{equation}
where $X_k$ is the kinetic energy at horizon crossing and $N_k$ is the remaining number of e-foldings of the ekpyrotic phase after $k$ mode exits horizon. (Factors of $p$ are neglected for the purpose of these estimates.)
Thus
\begin{equation} \label{eq:REk-end}
\Big. \mathcal{R}_k \Big| _{\text{ek-end}} \approx \mathcal{R}_k^{\text{const}} \approx \frac{C_1}{\sqrt{k}} \;
\gg \; \mathcal{R}_k^\text{int} \approx \frac{C_2 \sqrt{k}}{\sqrt{X_{\text{ek-end}}}}.
\end{equation}
Unfortunately, $\mathcal{R}_k^{\text{const}}$ has a blue spectrum ($P_\mathcal{R} \propto k^3 \vert \mathcal{R}_k \vert ^2 \propto k^2$), and $\mathcal{R}_k^\text{int}$ is bluer still ($P_\mathcal{R} \propto k^4$), inconsistent with the scale-invariant spectrum observed in CMB.

Following \cite{Buchbinder:2007ad}, we consider the entropic mechanism \cite{Lehners:2007ac} for generating a scale-invariant spectrum. By introducing two scalar fields $\phi_1$ and $\phi_2$ that undergo an ekpyrotic phase simultaneously, we have an extra degree of freedom that can source the curvature perturbation. Fluctuations in the fields are decomposed into an adiabatic mode $\delta \sigma$ along their mean trajectory and an entropic mode $\delta s$ perpendicular to it. Both fields obtain scale-invariant fluctuations during the ekpyrotic phase, but they source the curvature perturbation $\mathcal{R}$ differently \cite{Gordon:2000hv}, through the equation
\begin{equation} \label{eq:conversion}
\dot{\mathcal{R}} = \dot{\mathcal{R}}^{(\sigma)} - (1 + c_s^2) \frac{H}{\dot{\sigma}} \, \dot{\theta} \delta s.
\end{equation}
The first term represents the adiabatic contribution $\mathcal{R}^{(\sigma)}$ to the curvature perturbation, which is the same as in the single field case. The second term represents the entropic contribution $\mathcal{R}^{(s)}$, where $\theta$ is the angular direction of the trajectory in the $(\phi_1, \, \phi_2)$ plane. It is assumed \cite{Buchbinder:2007ad} that the trajectory is nearly straight except at the end of the ekpyrotic phase, where it undergoes a sharp bend and renders $\dot{\theta}$ temporarily non-zero. According to Eq.~(\ref{eq:conversion}), this causes the entropic perturbation $\delta s$ to convert almost instantaneously into a scale-invariant curvature perturbation $\mathcal{R}^{(s)}$, which remains constant on superhorizon scales afterwards.

Thus, the total curvature perturbation can be decomposed as
\begin{equation}
\mathcal{R}^\text{tot}= \mathcal{R}^{(s)}+\mathcal{R}^{(\sigma)} \approx \mathcal{R}^{(s)}+\mathcal{R}^{(\sigma,\text{const})}+ \mathcal{R}^{(\sigma,\text{int})},
\end{equation}
where as before the adiabatic contribution is divided into a constant term that is blue and an integral term that is bluer.
For the modes that exited horizon during the ekpyrotic phase, the constant term $\mathcal{R}^{(\sigma,\text{const})}$ is subdominant compared to the scale-invariant contribution $\mathcal{R}^{(s)}$, and the integral term $\mathcal{R}^{(\sigma,\text{int})}$ is {\it sub-subdominant},
\begin{align}
& \left| \frac{\mathcal{R}^{(\sigma,\text{const})}}{\mathcal{R}^{(s)}} \right|
\sim \frac{k}{(aH)_{\text{ek-end}}}
\sim \sqrt{ \frac{X_k}{X_{\rm ek-end}} }, \\
& \left| \frac{\mathcal{R}^{(\sigma,\text{int})}}{\mathcal{R}^{(s)}} \right|
\sim \Big( \frac{k}{(aH)_{\text{ek-end}}} \Big)^2
\sim \frac{X_k}{X_{\rm ek-end}}. \label{eq:REntro}
\end{align}  

The critical stage occurs towards the end of the kinetic phase after the conversion.
While $\mathcal{R}^{(s)}$ and $\mathcal{R}^{(\sigma,\text{const})}$ remain constant, $\mathcal{R}^{(\sigma,\text{int})}$ grows exponentially as $w \rightarrow -1$. The growth can be seen by rewriting Eq.~(\ref{eq:R0}) as
\begin{equation}
\mathcal{R}_k^{(\sigma,\text{int})} \approx C_2 \sqrt{k} \int \frac{c_s^2}{a^3} \frac{H^2}{-\dot{H}} dt \approx C_2 \sqrt{k} \int \frac{c_s^2}{a^3} \frac{2}{3(1+w)} \frac{dw}{\dot{w}}.
\end{equation}
Integrating $w$ from $w \gg 1$ at the end of the ekpyrotic phase to $w \approx -1$ \emph{just before} the non-linearity in $P(X)$ become significant, the dominant part of the integral comes from near $w \approx -1$.  Using Eq.~(\ref{eq:wdot}) and neglecting the second term as $w$ approaches
$-1$, we obtain
\begin{align} \label{eq:Rw-1}
\Big. \mathcal{R}_k^{(\sigma,\text{int})} \Big| ^{w \sim -1} &\approx C_2 \sqrt{k} \int^{\sim -1} \frac{c_s^2}{a^3} \frac{1}{3 H \big( \tfrac{-V_{,\phi}}{3 H^2} \big)} \frac{dw}{\sqrt{3 (1+w)^3}} \nonumber \\
&\approx \Big. C_2 \sqrt{k} \, \tfrac{c_s^2}{a^3} \big( \tfrac{2 H^2}{-V_{,\phi}} \big) \frac{1}{\sqrt{3(1+w) H^2}} \Big|^{w \sim -1} \nonumber \\
&\approx \Big. C_2 \sqrt{k} \, \frac{1}{\sqrt{X}} \Big| ^{X \sim X_{\text{c}}},
\end{align}
where we have neglected some finite factors that are almost constant during the rapid decrease in $w$, and used relation (\ref{eq:scaling}) since $P(X)$ is still linear.
The integral term has grown exponentially compared to its value at the end of the ekpyrotic phase from Eq.~(\ref{eq:REk-end}),
\begin{equation} \label{eq:Rw-1toEnd}
\left| \frac{\mathcal{R}^{(\sigma,\text{int})}_{w \sim -1}}{\mathcal{R}^{(\sigma,\text{int})}_{\text{ek-end}}} \right|
\approx \sqrt{\frac{X_\text{ek-end}}{X_\text{c}}}.
\end{equation}
Hence, from Eq.~(\ref{eq:REntro}), the ratio of the integral term to the scale-invariant entropic contribution can be expressed as
\begin{equation} \label{eq:Rw-1toEntro}
\left| \frac{\mathcal{R}^{(\sigma,\text{int})}_{w \sim -1}}{\mathcal{R}^{(s)}} \right|
\approx \frac{X_k}{X_\text{ek-end}} \, \sqrt{\frac{X_\text{ek-end}}{X_\text{c}}}.
\end{equation}

This ratio determines whether the adiabatic contribution can catch up with the entropic contribution and dominate the curvature perturbation. There is a competition between two factors:
(i) an exponential suppression from (\ref{eq:REntro}) that depends on how much $X$ {\it increases} after horizon exit during the ekpyrotic phase, \emph{i.e.} $X_k / X_\text{ek-end} \approx e^{-2 N_k}$ as in Eq.~(\ref{eq:Nk});
and (ii) an exponential amplification from (\ref{eq:Rw-1toEnd}) that depends on how much $X$ {\it decreases} in the kinetic phase, \emph{i.e.} $X_\text{ek-end} / X_\text{c} > X_\text{ek-end} / X_\text{ek-beg} \equiv e^{2 N_\text{tot}}$ according to Eq.~(\ref{eq:efolds}). Therefore we find, in total,
\begin{equation} \label{eq:Ntot-2Nk}
\left| \frac{\mathcal{R}^{(\sigma,\text{int})}_{w \sim -1}}{\mathcal{R}^{(s)}} \right|
\gtrsim e^{N_\text{tot} - 2 N_k}.
\end{equation}
The inequality boils down to the fact that the increase of the kinetic energy $X$ in the ekpyrotic phase is less than the decrease in the kinetic phase, hence amplification wins: This ratio is much greater than unity for modes with $N_k < N_\text{tot}/2$, which includes all modes within our observable horizon. That is, the integral adiabatic contribution, which has a doubly blue spectrum and was sub-subdominant at the end of the ekpyrotic phase, has grown to overwhelm the scale-invariant entropic contribution on relevant scales by the time the bouncing phase begins.

Thus, the dominantly blue spectrum will be carried into the expanding phase, in contradiction to current observations. Moreover, as Eq.~(\ref{eq:Ntot-2Nk}) dictates, for a wide range of $k$ modes the perturbation amplitudes have grown non-linear, to the extent that perturbation theory may break down even before the bounce.
 
As captured in Fig.~\ref{fig:kinetic}, the problem with the non-singular bounce arises from the generic requirement that in the kinetic phase the kinetic energy density must decrease by more than it has increased in the ekpyrotic phase, in order to trigger the ghost-condensate phase and violate the NEC. Hence, it is difficult to avoid this problem in a ghost-condensate bouncing model with a canonical (linear $P(X)$) ekpyrotic phase that does not involve extreme fine-tuning. Although the hope had been that a non-singular bounce would make ekpyrotic models simpler than the singular case by not having to consider quantum gravity effects, we are forced to conclude that this scenario leads to the exponential growth of adiabatic perturbations that spoil the spectrum of density perturbations.

As a possible remedy, the growth of curvature perturbations may be moderated if the ekpyrotic phase is realized with some non-linear form of $P(X, \phi)$, \emph{e.g.} in \cite{Creminelli:2007aq, withKhoury}. This kind of non-linearity is typically associated with quantum gravity effects, as occurs naturally in singular bounces. Hence, the simple non-singular bounce may fail for the reasons described in this paper, but it remains possible that a non-linear realization \cite{withKhoury}, or a singular bounce, can produce an observationally acceptable, nearly scale-invariant spectrum of density perturbations.

\begin{acknowledgments}

We thank Justin Khoury, Andrew Tolley, and Neil Turok for useful discussions.  The work is supported in part by Department of Energy Grant DE-FG02-91ER40671.

\end{acknowledgments}

\end{document}